%% file: paper.tex
\documentclass[10pt,conference]{IEEEtran}
\IEEEoverridecommandlockouts

\usepackage{cite}
\usepackage{amsmath,amssymb,amsfonts}
\usepackage{algorithm,algorithmic}
\usepackage{graphicx}
\usepackage{textcomp}
\usepackage{xcolor}
\usepackage[acronym]{glossaries}
\usepackage{blindtext}
\def\BibTeX{{\rm B\kern-.05em{\sc i\kern-.025em b}\kern-.08em
    T\kern-.1667em\lower.7ex\hbox{E}\kern-.125emX}}
\usepackage{booktabs}
\usepackage{bm}
\usepackage{import}
\usepackage{physics}
\usepackage{url}

\definecolor{color_vanilla}{rgb}{1.0, 0.5, 0.0}
\definecolor{color_natural}{rgb}{0.0, 0.5, 0.0}
\definecolor{color_natural_reg}{rgb}{0.0, 0.0, 1.0}
\definecolor{color_path_0}{rgb}{1.0, 0.15, 0.0}
\definecolor{color_path_1}{rgb}{0.0, 0.35, 0.0}
\definecolor{color_path_2}{rgb}{1.0, 0.65, 0.0}
\definecolor{color_path_3}{rgb}{0.0, 0.65, 0.0}

\definecolor{color_simulation}{rgb}{1.0, 0.0, 0.0}
\definecolor{color_hardware}{rgb}{0.0, 0.0, 1.0}

\usepackage{pgfplots}
\pgfplotsset{compat=1.14}
\usepgfplotslibrary{fillbetween}

\usepackage[raggedright]{subfigure}

\usepackage[capitalize,noabbrev]{cleveref}

\input{figures/data/training_trajectory}
\newcommand\copyrighttext{%
  \footnotesize \textcopyright 2023 IEEE. Personal use of this material is permitted. Permission from IEEE must be obtained for all other uses, in any current or future media, including reprinting/republishing this material for advertising or promotional purposes, creating new collective works, for resale or redistribution to servers or lists, or reuse of any copyrighted component of this work in other works.}
\newcommand\copyrightnotice{%
\begin{tikzpicture}[remember picture,overlay]
\node[anchor=south,yshift=10pt] at (current page.south) {\fbox{\parbox{\dimexpr\textwidth-\fboxsep-\fboxrule\relax}{\copyrighttext}}};
\end{tikzpicture}%
}

\begin{document}

\input{glossary.tex}

\title{Quantum Natural Policy Gradients:\\Towards Sample-Efficient Reinforcement Learning
    \thanks{The research was supported by the Bavarian Ministry of Economic Affairs, Regional Development and Energy with funds from the Hightech Agenda Bayern via the project BayQS and by the Bavarian Ministry for Economic Affairs, Infrastructure, Transport and Technology through the Center for Analytics-Data-Applications (ADA) within the framework of “BAYERN DIGITAL II”.\\
    M. Hartmann acknowledges support by the European Union’s Horizon 2020 research and innovation programme under grant agreement No 828826 “Quromorphic” and the Munich Quantum Valley, which is supported by the Bavarian state government with funds from
the Hightech Agenda Bayern Plus.\\
   Corresponding author: nico.meyer@iis.fraunhofer.de}
}


\author{
    \IEEEauthorblockN{
        Nico Meyer\IEEEauthorrefmark{1}\IEEEauthorrefmark{2},
        Daniel D. Scherer\IEEEauthorrefmark{1},
        Axel Plinge\IEEEauthorrefmark{1},
        Christopher Mutschler\IEEEauthorrefmark{1},
        Michael J. Hartmann\IEEEauthorrefmark{2}
    }
    \IEEEauthorblockA{
        \IEEEauthorrefmark{1}Fraunhofer IIS, Fraunhofer Institute for Integrated Circuits IIS, Nürnberg, Germany\\
        \IEEEauthorrefmark{2}Friedrich-Alexander University Erlangen-Nürnberg (FAU), Department of Physics, Erlangen, Germany
    }
}

\maketitle

\copyrightnotice

\begin{abstract}
    Reinforcement learning is a growing field in AI with a lot of potential. Intelligent behavior is learned automatically through trial and error in interaction with the environment. However, this learning process is often costly. Using \glsentrylongpl{vqc} as function approximators potentially can reduce this cost. 
    In order to implement this, we propose the \gls{qnpg} algorithm -- a second-order gradient-based routine that takes advantage of an efficient approximation of the quantum Fisher information matrix. We experimentally demonstrate that \gls{qnpg} outperforms first-order based training on different Contextual Bandits environments regarding convergence speed and stability and moreover reduces the sample complexity. Furthermore, we provide evidence for the practical feasibility of our approach by training on a $12$-qubit hardware device.
\end{abstract}


\glsresetall
\section{Introduction}
One critical technical factor in both classical and quantum \gls{rl} is the sample complexity, as interaction with the environment is potentially costly. Enhancing \gls{rl} with \glspl{vqc} as function approximators is a potential approach to reduce this cost utilizing the current noisy quantum hardware. 

The concept can be leveraged as a platform for \gls{qml}~\cite{Benedetti_2019}, which provides a provable quantum advantage for specific problems~\cite{Liu_2021,Sweke_2021}. Concrete realizations typically combine a \gls{vqc} with a classical training routine. This approach is believed to have some robustness to the (currently) inevitable hardware noise~\cite{Sharma_2020,Fontana_2021}. \gls{vqc} parameter updates can be computed using first-order gradients~\cite{Mitarai_2018}.

\Gls{qrl}~\cite{Meyer_2022a} aims at enhancing classical reinforcement learning~\cite{Sutton_2018} with quantum computing. \Gls{nisq}-compatible instances of \gls{qrl} employ \gls{vqc}-based function approximators for quantum Q-learning~\cite{Jerbi_2021} and \gls{qpg}~\cite{Jerbi_2021} approaches.

A concern for both \gls{qml} and \gls{qrl} is the trainability of the \gls{vqc}, and the associated sample complexity~\cite{Lattimore_2013}, i.e.,\ the required interactions with the environment. 
One can include second-order terms for more targeted parameter update~\cite{Amari_1998,Martens_2020} -- at the expense of circuit evaluation overhead.

\begin{figure}
    \centering
    \def\svgwidth{0.98\linewidth}
    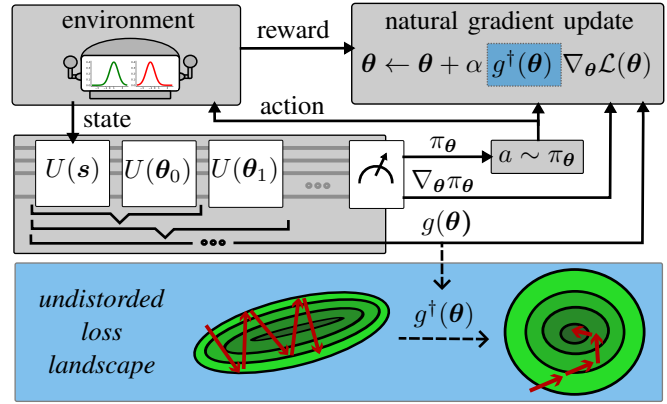
    \caption{Proposed method: The update with the first-order gradient $\nabla_{\bm{\theta}} \pi_{\bm{\theta}}$ is extended with a second-order term $g(\bm{\theta})$. This defines a \emph{(quantum) natural gradient} approach, which aims for training in a partially undistorted neighborhood of the parameter space -- improving convergence behavior.}
    \label{fig:figure_one}
    \vspace*{-4mm}
\end{figure}

\textbf{Contribution.} We propose a second-order extension
\footnote{Additional assumptions are necessary to constitute a formal approximate second-order technique~\cite{Stokes_2020}. However, this interpretation offers a good intuition and is therefore used throughout the paper.} 
to the \gls{qpg} algorithm~\cite{Jerbi_2021}. The idea of training in an undistorted neighborhood of the parameter space via the \gls{fim} is discussed in \cref{subsec:geometry_parameter_space}. We describe an efficient approximation of the quantum \gls{fim} in \cref{subsec:qfim} and propose a novel \gls{qnpg} algorithm in \cref{subsec:qnpg} (sketched in \cref{fig:figure_one}). We present empirical evidence that the \gls{qnpg} algorithm outperforms its first-order based counterpart on a proof-of-concept $1$-qubit \texttt{ContextualBandits} setup (\cref{subsec:poc}), but also performs well on a $12$-qubit hardware system (\cref{subsec:upscaled}).


\textbf{Related Work.} This work is based on a \gls{vqc}-based \gls{qpg} algorithm~\cite{Jerbi_2021} with classical post-processing~\cite{Meyer_2023}. Other extensions consider quantum-accessible environments~\cite{Jerbi_2022} and analyze the impact of hardware noise~\cite{Skolik_2023}. Our algorithm employs techniques for a block-diagonal approximation of the quantum \gls{fim}~\cite{Stokes_2020} and is inspired by classical natural policy gradients~\cite{Kakade_2001}. Quantum natural gradient techniques have also been investigated for the broader context of \gls{qml}~\cite{Haug_2021,Thanasilp_2021,Yamamoto_2019}.

\section{\label{sec:method}Method}

Time-dependent decision-making tasks in the presence of uncertainty can be addressed by \gls{rl}, where data is generated by an agent's interaction with the environment. This can be framed as a five-element \gls{mdp} ($\mathcal{S}, \mathcal{A}, R, T, \gamma)$, where $\mathcal{S}$ is a set of states, $\mathcal{A}$ describes the action set, a scalar reward function $R$, environment dynamics $T$, and discount factor $0 \leq \gamma \leq 1$. At each timestep $t$, the agent observes the environment state $\bm{s}_t$, and selects an action $a_t$ following the policy $\pi: \mathcal{S} \times \mathcal{A} \mapsto [0,1]$. The selected action is executed, and -- following its dynamics $T$ -- the environment returns a reward $r_t$, and transitions to the next state $s_{t+1}$. Good performance usually requires updating the policy to maximize the (discounted) return $G_t \gets \sum_{t'=t}^{H-1} \gamma^{t'-t}r_{t'}$ over some finite horizon $H < \infty$~\cite{Sutton_2018}.

To allow for a flexible modeling and updating of the policy, a parameterized function approximator $\pi_{\bm{\theta}}$ is used. The REINFORCE algorithm~\cite{Sutton_1999} -- referred to as \emph{vanilla policy gradients} -- allows to update the policy via gradient ascent steps $\bm{\theta} \gets \bm{\theta} + \alpha \nabla_{\bm{\theta}} \mathcal{L}(\bm{\theta})$. Here, $\alpha$ denotes the learning rate, and the gradient of the scalar performance measure $\mathcal{L}(\bm{\theta})$ is given by the policy gradient theorem as $\nabla_{\bm{\theta}} \mathcal{L}(\bm{\theta}) = \mathbb{E}_{\pi_{\bm{\theta}}} [ \sum_{t=0}^{H-1} \nabla_{\bm{\theta}} \ln \pi_{\bm{\theta}} (a_t | \bm{s}_t) \cdot G_t ] $~\cite{Sutton_1999}.

One promising type of parameterized function approximators -- besides the frequently used (deep) neural networks -- are \glspl{vqc}. This work starts from a \gls{qpg} algorithm~\cite{Jerbi_2021}, where measurements on the prepared quantum state are performed in the computational basis. Subsequent classical post-processing allows estimating the policy with $K$ shots as
\begin{equation}
    \pi_{\bm{\theta}} (a | \bm{s}) \approx \frac{1}{K} \sum_{k=0}^{K-1} \delta_{f_{\mathcal{C}}(\bm{b}^{(k)}) = a}
\end{equation}
where $\delta$ is an indicator function and $\bm{b}^{(k)}$ denotes the bitstring measured in the $k$-th shot~\cite{Meyer_2023}. As the experiments in this work are restricted to two actions, the post-processing function is selected as $f_{\mathcal{C}}(\bm{b}) = \bigoplus_{i=0}^{n-1} b_i$, with $n$ the number of qubits, and $b_i$ the $i$-th digit of the binary expansion of $\bm{b}$.

\subsection{\label{subsec:geometry_parameter_space}Capturing the Geometry of Parameter Space}

The vanilla \gls{qpg} algorithm performs the update of the parameters in the direction of the first-order gradient:
\begin{equation}
    \Delta \bm{\theta} = \nabla_{\bm{\theta}} \mathcal{L}(\bm{\theta}).
    \label{eq:vanilla_update}
\end{equation}
Contrarily, natural gradients are a second-order technique, and thus take the local curvature of parameter space into account.

Training the policy can be interpreted as minimizing distances between probability distributions. The vanilla update term from \cref{eq:vanilla_update} is closely tied to the Euclidean geometry, which is a sub-optimal choice in general. This gives rise to the idea to perform optimization directly on the statistical manifold defined by the parameters using the \gls{fim}:
\begin{equation}
    F(\bm{\theta}) = \mathbb{E}_{x \sim \pi_{\theta}(x)} \left[ \nabla_{\bm{\theta}} \ln \pi_{\bm{\theta}}(x) \nabla_{\bm{\theta}} \ln \pi_{\bm{\theta}}(x) ^ T \right].
\end{equation}
It locally approximates the Kullback-Leibler divergence~\cite{Kullback_1951}, i.e., describes the curvature of the parameter space around $\bm{\theta}$~\cite{Amari_1998}.
The inverse \gls{fim} can be used to perform updates in an undistorted neighborhood, as also sketched in \cref{fig:figure_one}. The \emph{natural gradient} update therefore is:
\begin{equation}
    \label{eq:natural_update}
    \Delta \bm{\theta} = \alpha F^{-1}(\bm{\theta}) \nabla_{\bm{\theta}} \mathcal{L} (\bm{\theta})
\end{equation}
This modified update rule offers advantages like invariance w.r.t.\ parameterization and stronger convergence guarantees than the vanilla approach~\cite{Martens_2020}. 

\subsection{\label{subsec:qfim}Quantum Generalization: Fubini-Study Metric Tensor}

In principle it would be possible to abstract the policy from the quantum model and use the classical \gls{fim} to define second-order gradient updates. However, this approach cannot capture the geometry underlying the quantum states produced by the \gls{vqc}, and is therefore missing the target. The Fubini-Study metric tensor -- also referred to as \emph{quantum \gls{fim}} -- provides a generalization of the \gls{fim} to the quantum case~\cite{Cheng_2010}. While the exact computation of this tensor on quantum hardware is not feasible in general, a diagonal or block-diagonal approximation can be obtained quite efficiently~\cite{Stokes_2020}.

\subsection{\label{subsec:qnpg}Quantum Natural Policy Gradients Algorithm}

We can formulate the objective for the \gls{rl} setup in resemblance with Eq. (9) of the work by Stokes et al.~\cite{Stokes_2020} as:
\begin{equation}
    \underset{\bm{\theta}}{\max} ~ \mathcal{L}(\bm{\theta}) = \underset{\bm{\theta}}{\max} ~ \mathbb{E}_{s \sim \mathcal{S}} \left[ \sum\nolimits_{a \in {\mathcal{A}}} \pi_{\bm{\theta}} (a | s) \cdot R(s,a) \right]
\end{equation}
This allows to incorporate the quantum \gls{fim} and formulate the \emph{quantum natural gradient} update rule:
\begin{equation}
    \label{eq:quantum_natural_update}
    \Delta \bm{\theta} = g^{\dagger}(\bm{\theta}) \nabla_{\bm{\theta}} \mathcal{L} (\bm{\theta})
\end{equation}
It has to be noted, that this update rule -- in contrast to the classical natural gradient for neural networks -- only describes the curvature of the objective up to some error. However, practical benefits compared to purely first-order based methods have been demonstrated for \gls{qml} methods~\cite{Yamamoto_2019,Stokes_2020}. We extent this analysis with our results on \gls{qrl} in \cref{sec:experiments}.

With the update rule from \cref{eq:quantum_natural_update} we formalize the \gls{qnpg} routine in \cref{alg:qnpg}. It is inspired by comparable classical approaches~\cite{Kakade_2001} and constitutes an extension of the typical Monte Carlo REINFORCE algorithm with second-order gradients. The first modification is line $7$, where we estimate a diagonal or block-diagonal approximation of the quantum \gls{fim}, as discussed in \cref{subsec:qfim}. 

\input{figures/single_qubit}

The second modification is line $8$ of \cref{alg:qnpg}, where we compute the pseudoinverse of the approximated metric tensor $g(\bm{\theta})$. For this one has to take into account, that the matrix of size $\abs{\bm{\theta}} \times \abs{\bm{\theta}}$ not necessarily has full rank. Let the combined first and second-order update (with dropped dependence on $t$) be defined as $\bm{\eta} := g^{\dagger}(\bm{\theta}) \nabla_{\bm{\theta}} \mathcal{L}(\bm{\theta})$, which is equivalent to solving for $\bm{\eta}_t$ in $g(\bm{\theta}) \bm{\eta} = \nabla_{\bm{\theta}} \mathcal{L}(\bm{\theta})$. The least squares solution~\cite{Lopatnikova_2021} for this problem is given by
\begin{equation}
    \hat{\bm{\eta}} = \underset{\bm{\eta}}{\mathrm{arg min}} \left\| g(\bm{\theta}) \bm{\eta} - \nabla_{\bm{\theta}} \mathcal{L} (\bm{\theta}) \right\|^2_2,
\end{equation}
which can be solved by basically any regression method. To ensure a more robust behavior, this formulation can be extended with regularization, i.e., ridge regression~\cite{Malago_2013}:
\begin{equation}
        \hat{\bm{\eta}}_{\xi} = \underset{\bm{\eta}}{\mathrm{arg min}} \left\| g(\bm{\theta}) \bm{\eta} - \nabla_{\bm{\theta}} \mathcal{L} (\bm{\theta}) \right\|^2_2 + \xi \left\| \bm{\eta} \right\|^2_2,
\end{equation}
where $\xi \in \mathbb{R}^+$ determines the influence of the penalty term. This regularization technique punishes large values in $\bm{\eta}$, which stabilizes the natural gradient update, in case the problem is ill-posed. For this work, we worked with the native \texttt{Qiskit} implementation~\cite{Qiskit_2023} to determine the hyperparameter $\xi$~\cite{Cultrera_2020}.

\begin{algorithm}[t]
    \caption{\label{alg:qnpg}Quantum Natural Policy Gradients (QNPG)}
    \begin{algorithmic}[1]
        \renewcommand{\algorithmicrequire}{\textbf{Input:}}
        \renewcommand{\algorithmicensure}{\textbf{Output:}}
        \REQUIRE initial policy $\pi_{\bm{\theta}}$, batch size $B$, discount factor $\gamma$, learning rate $\alpha$, termination condition
        \ENSURE policy $\pi_{\bm{\theta}_*}$ trained to maximize long term reward
        \WHILE{termination condition not satisfied}
            \STATE generate $B$ trajectories $\left[ s_0, a_0, r_0, s_1, a_1 \cdots \right]$ from $\pi_{\bm{\theta}}$
            \FOR{all trajectories $\tau$ in batch}
                \FOR{timestep $t$ in $0, \cdots, H-1$}
                    \STATE compute discounted returns $G_t \gets \sum_{t'=t}^{H-1} \gamma^{t'-t}r_{t'}$
                    \STATE sample first-order gradients $\nabla_{\bm{\theta}} \ln \pi_{\bm{\theta}} (a_t | \bm{s}_t)$
                    \STATE estimate Fubini-Study metric tensor $g(\bm{\theta})$
                    \STATE solve for $\bm{\eta}_t$ in $g(\bm{\theta}) \bm{\eta}_t = \nabla_{\bm{\theta}} \ln \pi_{\bm{\theta}} (a_t | \bm{s}_t)$
                \ENDFOR
            \ENDFOR
            \STATE compute batch average $\Delta \bm{\theta} \gets \frac{1}{B \cdot H} \sum_{\tau} \sum_{t=0}^{H-1} \bm{\eta}_t G_t$
            \STATE perform gradient ascent update $\bm{\theta} \gets \bm{\theta} + \alpha \Delta \bm{\theta}$
        \ENDWHILE
    \end{algorithmic}
\end{algorithm}

While \cref{alg:qnpg} is stated for a generic \gls{rl} problem, our experiments are conducted for the special case of one-step \texttt{ContextualBandits}, i.e., the horizon is $H=1$. The first-order gradients in line $6$ can be determined by applying the chain rule $\nabla_{\bm{\theta}} \ln \pi_{\bm{\theta}}(a | s) = \nabla_{\bm{\theta}} \pi_{\bm{\theta}}(a | s) / \pi_{\bm{\theta}}(a | s)$ and sampling parameter-shift gradients~\cite{Mitarai_2018}. This introduces the overhead of evaluating $2 \cdot \abs{\bm{\theta}}$ expectation values, which overshadows the additional complexity of approximating the quantum \gls{fim}. For a typical \gls{vqc} architecture, where each of the $n$ qubits is acted on with a parameterized rotation in a layer, this can be done with $\abs{\bm{\theta}} / n$ different circuits. 
While there are ways to avoid the size-dependent scaling of estimating first-order gradients via e.g. simultaneous perturbation stochastic approximation~\cite{Spall_1998}, similar techniques also exist for second-order gradients~\cite{Gacon_2021}. However, explicit consideration of this is out of the scope of this work, as we found the scaling of determining a (block-)diagonal approximation of the quantum \gls{fim} to be perfectly feasible for our purposes.

\input{figures/single_qubit_extended}

\section{\label{sec:experiments}Experiments}

We now demonstrate, that the proposed \gls{qnpg} algorithm empirically improves the convergence behavior -- as opposed to using the vanilla update rule -- for the \texttt{ContextualBandits} scenario~\cite{Sutton_2018}. We start with a proof-of-concept experiment in \cref{subsec:poc} and extend this to a $12$-qubit setup in \cref{subsec:upscaled}. Unless stated differently, we use the noiseless \texttt{QasmSimulator} of the \texttt{Qiskit} library, with $1024$ shots for estimating expectation values.

The state space of our \texttt{ContextualBandits} environments grows exponentially with the number of qubits in the employed \gls{vqc}, i.e., $\mathcal{S} = \left\{ 0, 1, \dots, 2^n-1 \right\}$. This allows for binary representation and encoding of the environment states $\bm{s} = s_0 s_1 \dots s_{n-1}$. Each state entails two actions $\mathcal{A} = \left\{ 0, 1 \right\}$, where the reward is either drawn from a Gaussian distribution $\mathcal{N}(\mu=-1, \sigma=1)$ or $\mathcal{N}(\mu=+1, \sigma=1)$. The task of the agent is therefore to identify the action which is associated with a mean value of $\mu=+1$ for each state individually. However, as the rewards $r_t$ are rather noisy, we consider the expected reward at timestep $t$ as a performance measure:
\begin{equation}
    \label{eq:expected_reward_timestep}
    \expval{r_t} = \pi_{\bm{\theta}}(a_{\text{opt}} | \bm{s}_t) - \pi_{\bm{\theta}} (\bar{a}_{\text{opt}} | \bm{s}_t),
\end{equation}
where $a_{\text{opt}}$ denotes the action that is optimal for state $\bm{s}_t$, and $\bar{a}_{\text{opt}}$ is the inverse choice. By averaging over the entire state space -- with equal weights, as each state is sampled uniformly at random -- we get the expected reward of the current policy
\begin{equation}
    \label{eq:expected_reward}
    \expval{r} = \frac{1}{2^n} \sum_{\bm{s} \in \left\{ 0, 1 \right\}^n} \pi_{\bm{\theta}}(a_{\text{opt}} | \bm{s}) - \pi_{\bm{\theta}} (\bar{a}_{\text{opt}} | \bm{s}).
\end{equation}
This metric is only used for tracking the training progress, while the agent only has access to the raw reward values $r_t$.

\subsection{\label{subsec:poc}Proof of Concept Demonstration on 1-Qubit System}

We consider a $2$-state \texttt{ContextualBandits} environment, where the optimal action is the same for both states. As the \gls{vqc} has only $2$ trainable parameters, we can visualize the expected reward from \cref{eq:expected_reward} over the entire periodic parameter space. While experiments are depicted for a learning rate of $\alpha=0.01$ and a single element per batch, similar results were observed for other hyperparameters.
The natural gradient technique (both regularized and non-regularized) clearly shows a faster convergence than the vanilla version. 


We provide detailed results for two specific regions of the parameter space in \cref{fig:single_qubit_extended}. The purely first-order based algorithm struggles when initialized in a distorted region of the parameter space. The performance oscillates until the agent is able to leave the distorted region after approx. $300$ episodes. In contrast, the natural gradient update enables traversing this region of the parameter space much faster. This also indicated a certain improvement in training stability, which has been of concern for \gls{vqc}-based \gls{qrl}~\cite{Franz_2022}. When initialized near the minimum, both agents can locate the optimum, but again the second-order version does so with fewer samples.

\subsection{\label{subsec:upscaled}Up-Scaling to a $12$-Qubit Hardware Device}

The $12$-qubit system in \cref{fig:multiple_qubits} allows working with a $2^{12}=4096$-state \texttt{ContextualBandits} environment, where the optimal action is given by the binary parity of the state, i.e., $a_{opt} = \bigoplus_{i=0}^{n-1} s_i$. This is encoded into the complex phase of the qubits, such that $s_i=0$ gives $\ket{R} = \frac{1}{\sqrt{2}} \left( \ket{0} + i \ket{1} \right)$ and $s_i=1$ produces $\ket{L} = \frac{1}{\sqrt{2}} \left( \ket{0} - i \ket{1} \right)$. An optimal parameter set for the $36$ $1$-qubit rotations is given in \cref{fig:multiple_qubits}. 
We initialize the individual parameters randomly in a $\mathcal{N}(\mu=0,\sigma=0.5)$-neighborhood of the optimal solution to speed up convergence.

\input{tables/hardware_test}

\input{figures/multiple_qubits}

Also in this scenario the (non-regularized) \gls{qnpg} algorithm outperforms the \gls{qpg} version in convergence speed. This -- admittedly small but still significant -- improvement is especially desirable in tasks, where sample complexity is a major concern. The overhead when using the \gls{qnpg} algorithm is negligible, i.e., $760$ instead of $730$ circuits have to be evaluated per batch, which is an increase of only approx. $4\%$.

Last but not least, we perform training on the $27$-qubit IBM Quantum system \texttt{ibmq\_ehningen}~\cite{IBMquantum_2021}. We employ a sub-graph of $12$ qubits with circular connectivity, which eliminates the overhead of transpiling the two-qubit gates. We exchange the parameter-shift gradients for an SPSA approximation~\cite{Spall_1998,Wiedmann_2023} with $10$ samples, which reduces the circuits for one batch from $720$ to $200$. Training took approx. $12$ hours on the quantum device, separated over $4$ \texttt{Qiskit Sessions} executed over the timeframe of several days. While the employed matrix-free measurement error mitigation~\cite{Nation_2021,Qiskit_2023} improved the results by approx. $60\%$, more advanced techniques did not lead to significant refinements. The performance clearly declines compared to the simulation. First, noise flattens the loss landscape and therefore also the magnitude of gradients~\cite{Wang_2021a, Wang_2021b}, which slows down convergence. Second, the noise level of the current quantum devices does not allow measuring (near-)optimal expectation values, even knowing the optimal parameter set. We support this claim by evaluating the policy in \cref{tab:hardware_test} for both, the learned and analytically optimal parameters. The learned parameters produce an expected reward of $0.574$, while the optimal ones only get to $0.568$. While the advantage is not significant enough to attribute it to the algorithm's capability of inherently dealing with the noise -- as often claimed for \glspl{vqc}~\cite{Moll_2018,Sharma_2020,Fontana_2021} -- it demonstrates the feasibility of the \gls{qnpg} approach on quantum hardware.

\glsresetall

\section{Conclusion}

In this work, we address the trainability and associated sample complexity of a \gls{vqc}-based \gls{qpg} method~\cite{Jerbi_2021,Meyer_2023}. We proposed the \gls{qnpg} algorithm, which extends the vanilla \gls{qpg} approach by second-order terms. Inspired by classical natural gradients, the pseudoinverse of the quantum \gls{fim} is incorporated into the update procedure. This allows for more targeted updates in a partially undistorted neighborhood of the parameter space.

There are theoretical guarantees for classical natural gradient algorithms~\cite{Kakade_2001,Amari_1998,Martens_2020} and practical benefits for \glsentrylong{qml}~\cite{Yamamoto_2019,Stokes_2020}. We extend upon this analysis and provide evidence for the efficiency of the proposed routine for \glsentrylong{qrl}. On \texttt{ContextualBandits} environments of increasing complexity we show, that \gls{qnpg} is superior to \gls{qpg} in terms of convergence speed and therefore sample efficiency. The overhead for approximating the quantum \gls{fim} is negligible compared to sampling first-order gradients. With the \gls{qnpg} algorithm we also train on a $12$-qubit hardware device.

Due to the gathered results, we claim that \gls{qnpg} improves upon the original \gls{qpg}. While there are disputes regarding the impact of quantum natural gradients on barren plateaus~\cite{Thanasilp_2021}, there is evidence that the problem gets at least mitigated~\cite{Haug_2021}. Altogether, this work is a proof-of-concept step toward improving the training procedure of \gls{vqc}-based \gls{qrl} models.

\section*{Acknowledgment}

We wish to thank G. Wellein for his administrative and technical support of this work. The authors gratefully acknowledge the scientific support and HPC resources provided by the Erlangen National High Performance Computing Center (NHR@FAU) of the Friedrich-Alexander-Universität Erlangen-Nürnberg (FAU). The hardware is funded by the German Research Foundation (DFG).

Access to the IBM Quantum Services was obtained through the IBM Quantum Hub at Fraunhofer. The views expressed are those of the authors, and do not reflect the official policy or position of IBM or the IBM Quantum team.


\section*{Code Availability}

An implementation to reproduce the main results of this paper with a routine to approximate the quantum \gls{fim} is available at \url{https://gitlab.com/NicoMeyer/qnpg}. Further information and data is available upon reasonable request.

\bibliographystyle{IEEEtran}
\bibliography{paper}

\end{document}

%% file: glossary.tex
\newacronym{vqc}{VQC}{variational quantum circuit}
\newacronym{vqa}{VQA}{variational quantum algorithm}
\newacronym{rl}{RL}{reinforcement learning}
\newacronym{qrl}{QRL}{quantum reinforcement learning}
\newacronym{ml}{ML}{machine learning}
\newacronym{qml}{QML}{quantum machine learning}
\newacronym{dnn}{DNN}{deep neural network}
\newacronym{qc}{QC}{quantum computing}
\newacronym{nisq}{NISQ}{noisy intermediate-scale quantum}
\newacronym{pg}{PG}{policy gradient}
\newacronym{qpg}{QPG}{quantum policy gradient}
\newacronym{qnpg}{QNPG}{quantum natural policy gradient}
\newacronym{mdp}{MDP}{Markov Decision Process}
\newacronym{spsa}{SPSA}{simultaneous perturbation stochastic approximations}
\newacronym{fim}{FIM}{Fisher information matrix}
\newacronym{pdf}{PDF}{probability density function}

%% file: figures/qnpg.pdf_tex
\begingroup%
  \makeatletter%
  \providecommand\color[2][]{%
    \errmessage{(Inkscape) Color is used for the text in Inkscape, but the package 'color.sty' is not loaded}%
    \renewcommand\color[2][]{}%
  }%
  \providecommand\transparent[1]{%
    \errmessage{(Inkscape) Transparency is used (non-zero) for the text in Inkscape, but the package 'transparent.sty' is not loaded}%
    \renewcommand\transparent[1]{}%
  }%
  \providecommand\rotatebox[2]{#2}%
  \newcommand*\fsize{\dimexpr\f@size pt\relax}%
  \newcommand*\lineheight[1]{\fontsize{\fsize}{#1\fsize}\selectfont}%
  \ifx\svgwidth\undefined%
    \setlength{\unitlength}{692.27487639bp}%
    \ifx\svgscale\undefined%
      \relax%
    \else%
      \setlength{\unitlength}{\unitlength * \real{\svgscale}}%
    \fi%
  \else%
    \setlength{\unitlength}{\svgwidth}%
  \fi%
  \global\let\svgwidth\undefined%
  \global\let\svgscale\undefined%
  \makeatother%
  \begin{picture}(1,0.61215944)%
    \lineheight{1}%
    \setlength\tabcolsep{0pt}%
    \put(0,0){\includegraphics[width=\unitlength,page=1]{qnpg.pdf}}%
    \put(0.07763723,0.5731535){\color[rgb]{0,0,0}\makebox(0,0)[lt]{\lineheight{1.25}\smash{\begin{tabular}[t]{l}environment\end{tabular}}}}%
    \put(0.1383913,0.14501228){\color[rgb]{0,0,0}\makebox(0,0)[t]{\lineheight{1.25}\smash{\begin{tabular}[t]{c}\textit{undistorded}\\\textit{loss}\\\textit{landscape}\end{tabular}}}}%
    \put(0.57205687,0.57224038){\color[rgb]{0,0,0}\makebox(0,0)[lt]{\lineheight{1.25}\smash{\begin{tabular}[t]{l}natural gradient update\end{tabular}}}}%
    \put(0.74603801,0.36809626){\color[rgb]{0,0,0}\makebox(0,0)[lt]{\lineheight{1.25}\smash{\begin{tabular}[t]{l}$a \sim \pi_{\boldsymbol{\theta}}$\end{tabular}}}}%
    \put(0.05151003,0.34439798){\color[rgb]{0,0,0}\makebox(0,0)[lt]{\lineheight{1.25}\smash{\begin{tabular}[t]{l}$U(\boldsymbol{s})$\end{tabular}}}}%
    \put(0.17209278,0.34335538){\color[rgb]{0,0,0}\makebox(0,0)[lt]{\lineheight{1.25}\smash{\begin{tabular}[t]{l}$U(\boldsymbol{\theta}_0)$\end{tabular}}}}%
    \put(0.30532261,0.34425269){\color[rgb]{0,0,0}\makebox(0,0)[lt]{\lineheight{1.25}\smash{\begin{tabular}[t]{l}$U(\boldsymbol{\theta}_1)$\end{tabular}}}}%
    \put(0.6400501,0.38983097){\color[rgb]{0,0,0}\makebox(0,0)[lt]{\lineheight{1.25}\smash{\begin{tabular}[t]{l}$\pi_{\boldsymbol{\theta}}$\end{tabular}}}}%
    \put(0.38058284,0.44202799){\color[rgb]{0,0,0}\makebox(0,0)[lt]{\lineheight{1.25}\smash{\begin{tabular}[t]{l}action\end{tabular}}}}%
    \put(0.10949031,0.4213416){\color[rgb]{0,0,0}\makebox(0,0)[lt]{\lineheight{1.25}\smash{\begin{tabular}[t]{l}state\end{tabular}}}}%
    \put(0.61212715,0.32797389){\color[rgb]{0,0,0}\makebox(0,0)[lt]{\lineheight{1.25}\smash{\begin{tabular}[t]{l}$\nabla_{\boldsymbol{\theta}} \pi_{\boldsymbol{\theta}}$\end{tabular}}}}%
    \put(0.62793017,0.26330725){\color[rgb]{0,0,0}\makebox(0,0)[lt]{\lineheight{1.25}\smash{\begin{tabular}[t]{l}$g(\boldsymbol{\theta)}$\end{tabular}}}}%
    \put(0.61626037,0.1223325){\color[rgb]{0,0,0}\makebox(0,0)[lt]{\lineheight{1.25}\smash{\begin{tabular}[t]{l}$g^{\dagger}(\boldsymbol{\theta})$\end{tabular}}}}%
    \put(0.53550655,0.50526711){\color[rgb]{0,0,0}\makebox(0,0)[lt]{\lineheight{1.25}\smash{\begin{tabular}[t]{l}$\boldsymbol{\theta} \gets \boldsymbol{\theta} + \alpha ~ g^{\dagger}(\boldsymbol{\theta}) ~ \nabla_{\bm{\theta}} \mathcal{L}(\boldsymbol{\theta})$\end{tabular}}}}%
    \put(0,0){\includegraphics[width=\unitlength,page=2]{qnpg.pdf}}%
    \put(0.37615861,0.55772912){\color[rgb]{0,0,0}\makebox(0,0)[lt]{\lineheight{1.25}\smash{\begin{tabular}[t]{l}reward\end{tabular}}}}%
    \put(0,0){\includegraphics[width=\unitlength,page=3]{qnpg.pdf}}%
  \end{picture}%
\endgroup%

%% file: figures/single_qubit.tex
\begin{figure*}[t]
    \pgfplotsset{colormap/jet}
    \centering
    \subfigure[VQC with two trainable parameters. Repeated encoding enhances expressivity, similar to data re-uploading~\cite{Perez_2020} and incremental data uploading~\cite{Periyasamy_2022}.]{
        \includegraphics[width=0.98\linewidth]{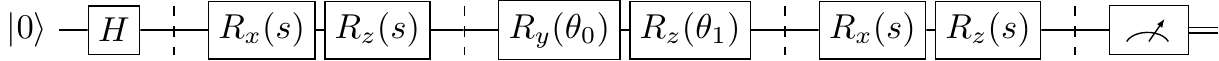}
    }
    \\
    \subfigure[Environment with $a_{\text{opt}}=0$.]{
        \def\svgwidth{0.23\linewidth}
        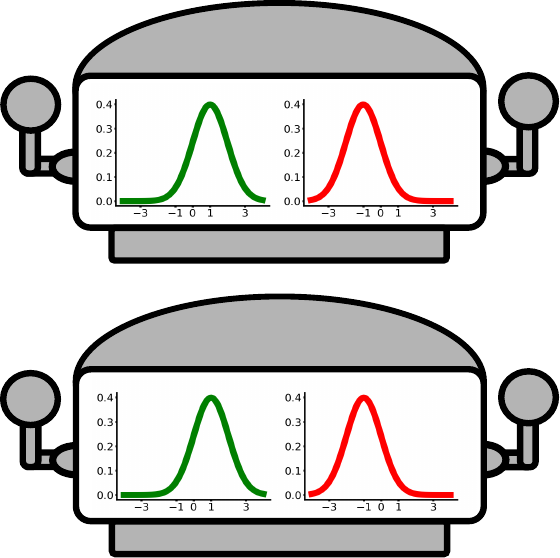    
    }
    \qquad
    \subfigure[Associated parameter landscape.]{
    \begin{tikzpicture}
        \centering
        \begin{axis}[
            name=plot1,
            xlabel=$\mathbf{\boldsymbol{\theta}_0}$,
            xlabel style={at={(0.78,-0.00)}},
            ylabel=$\mathbf{\boldsymbol{\theta}_1}$,
            ylabel style={at={(0.22,-0.00)}},
            zlabel=$\textbf{Expected Reward}$,
            zlabel style={at={(-0.15,0.5)}},
            xtick={-1.57, 0, 1.57},
            xticklabels={$-\frac{\pi}{2}$, , $\frac{\pi}{2}$},
            ytick={-1.57, 0, 1.57},
            zticklabel style={/pgf/number format/.cd,fixed,fixed zerofill,precision=1,},
            yticklabels={$-\frac{\pi}{2}$, , $\frac{\pi}{2}$},
            label style={font=\footnotesize},
            tick label style={font=\footnotesize},
            grid=both,
            view={135}{20},
            width=0.34\linewidth,
            zmin=-1.0,zmax=1.0
            ]
            \addplot3[surf, mesh/cols=45, faceted color = black] %
            	table[x=x,y=y,z=z,col sep=comma]{figures/data/env_unbalanced.csv};
        \end{axis}
    \end{tikzpicture}
    }
    \qquad
    \subfigure[Performance for $100$ random initializations.]{
        \begin{tikzpicture}
        \centering
        \begin{axis}[
            name=plot1,
            xlabel=$\textbf{Episode}$,
            ylabel=$\textbf{Average Expected Reward}$,
            xtick={0, 250, 500},
            ytick={0.0, 0.5, 1.0},
            yticklabel style={/pgf/number format/.cd,fixed,fixed zerofill,precision=1,},
            xmin=0,xmax=525,
            ymin=-0.1,ymax=1.1,
            label style={font=\footnotesize},
            tick label style={font=\footnotesize},
            grid=both,
            axis x line=bottom, axis y line=left,
            width=0.32\linewidth,
            legend style={/tikz/every even column/.append style={column sep=0.1cm, row sep=0.1cm},at={(0.97,0.38)},anchor=east,yshift=-5mm,font=\footnotesize},
        	no marks]
            ]
             \addplot+[draw=none,name path=A,no markers] %
            	table[x=episode,y=vanilla_var_neg,col sep=comma]{figures/data/perf_random_unbalanced.csv};
             \addplot+[draw=none,name path=B,no markers] %
            	table[x=episode,y=vanilla_var_pos,col sep=comma]{figures/data/perf_random_unbalanced.csv};
             \addplot[color_vanilla!40] fill between[of=A and B];

             \addplot+[draw=none,name path=E,no markers] %
            	table[x=episode,y=regnatural_var_neg,col sep=comma]{figures/data/perf_random_unbalanced.csv};
             \addplot+[draw=none,name path=F,no markers] %
            	table[x=episode,y=regnatural_var_pos,col sep=comma,on layer=main]{figures/data/perf_random_unbalanced.csv};
             \addplot[color_natural_reg!40] fill between[of=E and F];

             \addplot+[draw=none,name path=C,no markers] %
            	table[x=episode,y=natural_var_neg,col sep=comma]{figures/data/perf_random_unbalanced.csv};
             \addplot+[draw=none,name path=D,no markers] %
            	table[x=episode,y=natural_var_pos,col sep=comma]{figures/data/perf_random_unbalanced.csv};
             \addplot[color_natural!40] fill between[of=C and D];

             \addplot[line width=.7pt,solid,color=color_vanilla] %
            	table[x=episode,y=vanilla,col sep=comma]{figures/data/perf_random_unbalanced.csv};
             \addplot[line width=.7pt,solid,color=color_natural] %
            	table[x=episode,y=natural,col sep=comma]{figures/data/perf_random_unbalanced.csv};
             \addplot[line width=.7pt,solid,color=color_natural_reg] %
            	table[x=episode,y=regnatural,col sep=comma]{figures/data/perf_random_unbalanced.csv};

            \addplot[color=black, domain=0:520, line width=.5pt] {1.0};
            \addplot[dashed, color=black, domain=0:520, line width=.5pt] {0.0};

            \legend{,,,,,,,,,vanilla,natural,reg. nat.,,}

        \end{axis}
        \end{tikzpicture}
    }
    \caption{\label{fig:single_qubit}Experiment with a $1$-qubit circuit on a simple \texttt{ContextualBandits} environment with $\mathcal{S} = \lbrace 0, 1 \rbrace$ and $\mathcal{A} = \lbrace 0, 1 \rbrace$.
    }
\end{figure*}

%% file: figures/bandits_unbalanced.pdf_tex
\begingroup%
  \makeatletter%
  \providecommand\color[2][]{%
    \errmessage{(Inkscape) Color is used for the text in Inkscape, but the package 'color.sty' is not loaded}%
    \renewcommand\color[2][]{}%
  }%
  \providecommand\transparent[1]{%
    \errmessage{(Inkscape) Transparency is used (non-zero) for the text in Inkscape, but the package 'transparent.sty' is not loaded}%
    \renewcommand\transparent[1]{}%
  }%
  \providecommand\rotatebox[2]{#2}%
  \newcommand*\fsize{\dimexpr\f@size pt\relax}%
  \newcommand*\lineheight[1]{\fontsize{\fsize}{#1\fsize}\selectfont}%
  \ifx\svgwidth\undefined%
    \setlength{\unitlength}{161.01251077bp}%
    \ifx\svgscale\undefined%
      \relax%
    \else%
      \setlength{\unitlength}{\unitlength * \real{\svgscale}}%
    \fi%
  \else%
    \setlength{\unitlength}{\svgwidth}%
  \fi%
  \global\let\svgwidth\undefined%
  \global\let\svgscale\undefined%
  \makeatother%
  \begin{picture}(1,0.99674316)%
    \lineheight{1}%
    \setlength\tabcolsep{0pt}%
    \put(0,0){\includegraphics[width=\unitlength,page=1]{bandits_unbalanced.pdf}}%
    \put(0.03485757,0.25759841){\color[rgb]{0,0,0}\makebox(0,0)[lt]{\lineheight{1.25}\smash{\begin{tabular}[t]{l}0\end{tabular}}}}%
    \put(0.92607584,0.26399509){\color[rgb]{0,0,0}\makebox(0,0)[lt]{\lineheight{1.25}\smash{\begin{tabular}[t]{l}1\end{tabular}}}}%
    \put(0.92543569,0.79367829){\color[rgb]{0,0,0}\makebox(0,0)[lt]{\lineheight{1.25}\smash{\begin{tabular}[t]{l}1\end{tabular}}}}%
    \put(0.32371058,0.88945351){\color[rgb]{0,0,0}\makebox(0,0)[lt]{\lineheight{1.25}\smash{\begin{tabular}[t]{l}Bandit \#0\end{tabular}}}}%
    \put(0.32889909,0.36929188){\color[rgb]{0,0,0}\makebox(0,0)[lt]{\lineheight{1.25}\smash{\begin{tabular}[t]{l}Bandit \#1\end{tabular}}}}%
    \put(0.03394603,0.78399817){\color[rgb]{0,0,0}\makebox(0,0)[lt]{\lineheight{1.25}\smash{\begin{tabular}[t]{l}0\end{tabular}}}}%
  \end{picture}%
\endgroup%

%% file: figures/single_qubit_extended.tex
\begin{figure*}
    \centering
\subfigure[Initialized in distorted region.]{
        \begin{tikzpicture}
        \centering
        \begin{axis}[
            name=plot1,
            xlabel=$\textbf{Episode}$,
            ylabel=$\textbf{Average Expected Reward}$,
            xtick={0, 250, 500},
            ytick={-1.0, 0.0, 1.0},
            yticklabel style={/pgf/number format/.cd,fixed,fixed zerofill,precision=1,},
            xmin=0,xmax=525,
            ymin=-1.1,ymax=1.1,
            label style={font=\footnotesize},
            tick label style={font=\footnotesize},
            grid=both,
            axis x line=bottom, axis y line=left,
            width=0.27\linewidth,
            height=5.2cm,
            legend style={/tikz/every even column/.append style={column sep=0.1cm, row sep=0.1cm},at={(0.97,0.29)},anchor=east,yshift=-5mm,font=\footnotesize},
        	no marks]
            ]

             \addplot[line width=.3pt,solid,color=color_path_0!40] %
            	table[x=episode,y=first,col sep=comma]{figures/data/perf_flat.csv};
             \addplot[line width=.4pt,solid,color=color_path_0] %
            	table[x=episode,y=first_mean,col sep=comma]{figures/data/perf_flat.csv};

              \addplot[line width=.3pt,solid,color=color_path_1!40] %
            	table[x=episode,y=natural,col sep=comma]{figures/data/perf_flat.csv};
             \addplot[line width=.4pt,solid,color=color_path_1] %
            	table[x=episode,y=natural_mean,col sep=comma]{figures/data/perf_flat.csv};

            \addplot[color=black, domain=0:520, line width=.5pt] {1.0};
            \addplot[dashed, color=black, domain=0:520, line width=.5pt] {-0.3};

            \legend{,vanilla,,natural}

        \end{axis}
        \end{tikzpicture}
    }
    \qquad
    \subfigure[Associated trajectories ($\blacktriangle$ marks maximum).]{
    \includegraphics[width=0.30\linewidth]{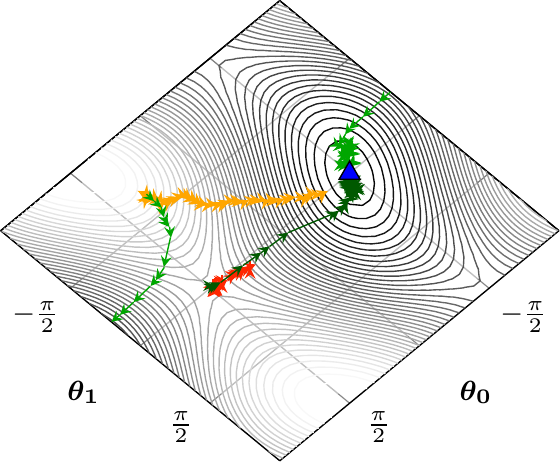}
    }
    \qquad
    \subfigure[Initialized near minimum.]{
        \begin{tikzpicture}
        \centering
        \begin{axis}[
            name=plot1,
            xlabel=$\textbf{Episode}$,
            ylabel=$\textbf{Average Expected Reward}$,
            xtick={0, 250, 500},
            ytick={-1.0, 0.0, 1.0},
            yticklabel style={/pgf/number format/.cd,fixed,fixed zerofill,precision=1,},
            xmin=0,xmax=525,
            ymin=-1.1,ymax=1.1,
            label style={font=\footnotesize},
            tick label style={font=\footnotesize},
            grid=both,
            axis x line=bottom, axis y line=left,
            width=0.27\linewidth,
            height=5.2cm,
            legend style={/tikz/every even column/.append style={column sep=0.1cm, row sep=0.1cm},at={(0.97,0.29)},anchor=east,yshift=-5mm,font=\footnotesize},
        	no marks]
            ]

            \addplot[line width=.3pt,solid,color=color_path_2!40] %
            	table[x=episode,y=first,col sep=comma]{figures/data/perf_min.csv};
             \addplot[line width=.4pt,solid,color=color_path_2] %
            	table[x=episode,y=first_mean,col sep=comma]{figures/data/perf_min.csv};

              \addplot[line width=.3pt,solid,color=color_path_3!40] %
            	table[x=episode,y=natural,col sep=comma]{figures/data/perf_min.csv};
             \addplot[line width=.4pt,solid,color=color_path_3] %
            	table[x=episode,y=natural_mean,col sep=comma]{figures/data/perf_min.csv};

            \addplot[color=black, domain=0:520, line width=.5pt] {1.0};
            \addplot[dashed, color=black, domain=0:520, line width=.5pt] {-0.78};

            \legend{,vanilla,,natural}

        \end{axis}
        \end{tikzpicture}
    }
    \caption{\label{fig:single_qubit_extended}
    With the vanilla and (non-regularized) natural gradient update technique $10$ agents were trained, and the trajectory of a random instance -- depicted with faded colors in (a) and (c) -- is tracked in the parameter landscape (b).}
\end{figure*}
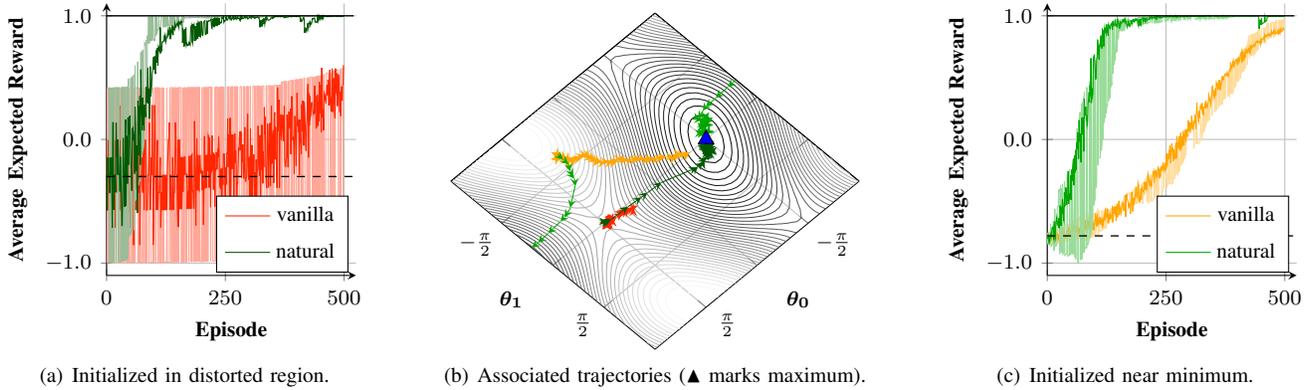

%% file: tables/hardware_test.tex
\begin{table}[t]
    \centering
    \caption{\label{tab:hardware_test}
    Percentage of states, for which the probability of selecting the optimal action is above the given thresholds, evaluated on \texttt{ibmq\_ehningen} hardware for full $4096$-element state space.}
        \begin{tabular}{ccccc}
            \toprule
            $\pi ( a_{\text{opt}} | \bm{s})$ & $\geq 0.95$ & $\geq 0.85$ & $\geq 0.75$ & $\geq 0.65$ \\
            \midrule
            trained parameters & $2.5\%$ & $52.8\%$ & $68.5\%$ & $79.3\%$ \\
            optimal parameters & $0.4\%$ & $61.0\%$ & $72.9\%$ & $75.0\%$ \\
            \bottomrule
            \toprule
           $\pi ( a_{\text{opt}} | \bm{s})$ & $\geq 0.55$ & $\geq 0.45$ & $\geq 0.35$ & $< 0.35$ \\
            \midrule
            trained parameters & $86.0\%$ & $95.8\%$ & $100.0\%$ & $100.0\%$ \\
            optimal parameters & $81.2\%$ & $89.1\%$ & $100.0\%$ & $100.0\%$ \\
            \bottomrule
        \end{tabular}
\end{table}

%% file: figures/multiple_qubits.tex
\begin{figure*}
    \pgfplotsset{colormap/jet}
    \centering
    \subfigure[VQC with phase-encoding $P(\theta) = \mathrm{diag}(1, e^{i \theta})$ of the $12$-dimensional binary environment states. Gray values denote an optimal parameter set for the $36$ rotations. Controlled-$X$ gates are applied to neighboring qubits with even-numbered (odd-numbered) ones as control in the first (second) entanglement layer.
    ]{
        \includegraphics[width=0.99\linewidth]{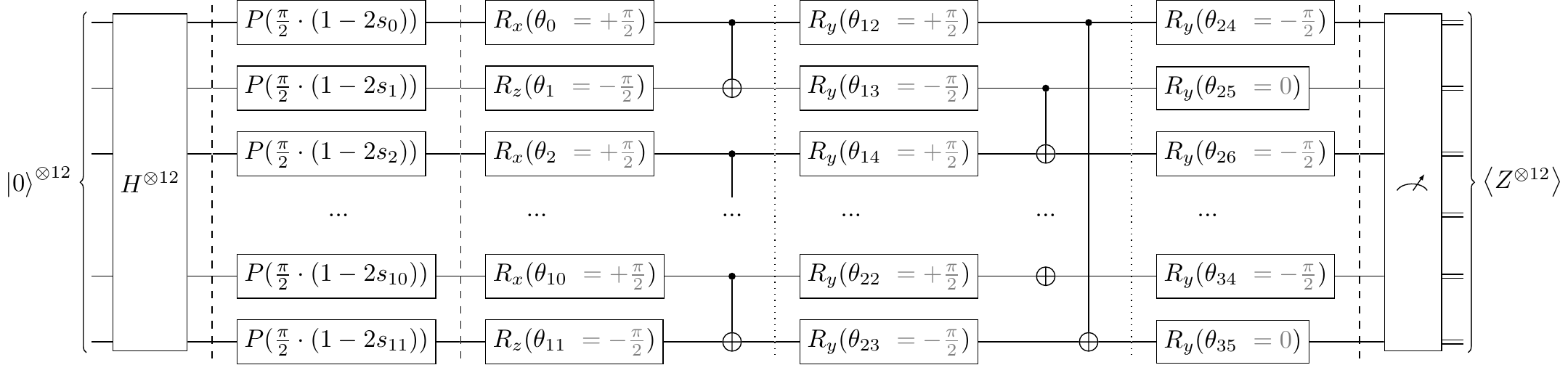}
    }
    \\
    \subfigure[$4096$-state environment.]{
        \def\svgwidth{0.23\linewidth}
        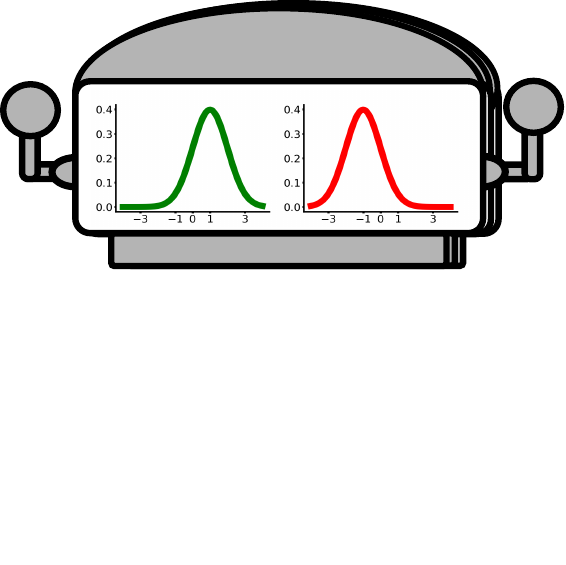    
    }
    \qquad
    \subfigure[Performance for $25$ random initializations.]{
                \begin{tikzpicture}
        \centering
        \begin{axis}[
            name=plot1,
            xlabel=$\textbf{Episode}$,
            ylabel=$\textbf{Average Expected Reward}$,
            xtick={0, 2500, 5000},
            xticklabels={$0$, $2.5$K, $5$K},
            ytick={0.0, 0.5, 1.0},
            yticklabel style={/pgf/number format/.cd,fixed,fixed zerofill,precision=1,},
            xmin=0,xmax=5025,
            ymin=-0.2,ymax=1.1,
            label style={font=\footnotesize},
            tick label style={font=\footnotesize},
            grid=both,
            axis x line=bottom, axis y line=left,
            width=0.32\linewidth,
            legend style={/tikz/every even column/.append style={column sep=0.1cm, row sep=0.1cm},at={(0.97,0.32)},anchor=east,yshift=-5mm,font=\footnotesize},
        	no marks]
            ]
             \addplot+[draw=none,name path=A,no markers] %
            	table[x=episode,y=vanilla_var_neg,col sep=comma]{figures/data/multiple_qubits_simulator.csv};
             \addplot+[draw=none,name path=B,no markers] %
            	table[x=episode,y=vanilla_var_pos,col sep=comma]{figures/data/multiple_qubits_simulator.csv};
             \addplot[color_vanilla!40] fill between[of=A and B];

             \addplot+[draw=none,name path=C,no markers] %
            	table[x=episode,y=natural_var_neg,col sep=comma]{figures/data/multiple_qubits_simulator.csv};
             \addplot+[draw=none,name path=D,no markers] %
            	table[x=episode,y=natural_var_pos,col sep=comma]{figures/data/multiple_qubits_simulator.csv};
             \addplot[color_natural!40] fill between[of=C and D];

             \addplot[line width=1pt,solid,color=color_vanilla] %
            	table[x=episode,y=vanilla_mean,col sep=comma]{figures/data/multiple_qubits_simulator.csv};
             \addplot[line width=1pt,solid,color=color_natural] %
            	table[x=episode,y=natural_mean,col sep=comma]{figures/data/multiple_qubits_simulator.csv};

            \addplot[color=black, domain=0:5020, line width=.5pt] {1.0};
            \addplot[dashed, color=black, domain=0:5020, line width=.5pt] {0.0};

            \legend{,,,,,,vanilla,natural}

        \end{axis}
        \end{tikzpicture}
    }
    \qquad
    \subfigure[Training on \texttt{ibmq\_ehningen} system.]{
                \begin{tikzpicture}
        \centering
        \begin{axis}[
            name=plot1,
            xlabel=$\textbf{Episode}$,
            ylabel=$\textbf{Expected Reward}$,
            xtick={0, 900, 1800},
            xticklabels={$0$, $900$, $1.8$K},
            ytick={0.0, 0.5, 1.0},
            yticklabel style={/pgf/number format/.cd,fixed,fixed zerofill,precision=1,},
            xmin=0,xmax=1825,
            ymin=-0.2,ymax=1.1,
            label style={font=\footnotesize},
            tick label style={font=\footnotesize},
            grid=both,
            axis x line=bottom, axis y line=left,
            width=0.32\linewidth,
            legend style={/tikz/every even column/.append style={column sep=0.1cm, row sep=0.1cm},at={(0.97,0.32)},anchor=east,yshift=-5mm,font=\footnotesize},
        	no marks]
            ]

             \addplot[line width=1pt,solid,color=color_simulation!40] %
            	table[x=episode,y=sim_batch,col sep=comma]{figures/data/multiple_qubits_hardware_batch.csv};
             \addplot[line width=1pt,solid,color=color_simulation] %
            	table[x=episode,y=sim_val,col sep=comma]{figures/data/multiple_qubits_hardware_validate.csv};

             \addplot[line width=1pt,solid,color=color_hardware!40] %
            	table[x=episode,y=hw_batch,col sep=comma]{figures/data/multiple_qubits_hardware_batch.csv};
             \addplot[line width=1pt,solid,color=color_hardware] %
            	table[x=episode,y=hw_val,col sep=comma]{figures/data/multiple_qubits_hardware_validate.csv};

            \addplot[color=black, domain=0:1820, line width=.5pt] {1.0};
            \addplot[dotted, color=blue, domain=0:1820, line width=1pt] {0.57};
            \addplot[dashed, color=black, domain=0:1820, line width=.5pt] {0.0};

            \legend{,simulation,,hardware}

        \end{axis}
        \end{tikzpicture}
    }
    \caption{\label{fig:multiple_qubits} Performances (in simulation) are compared for random parameter initializations (c). One complete training procedure with natural gradients is run on actual quantum hardware. Validation (dark curves) is conducted after each $10$ batches with a random selection of $256$ states to increase interpretability (d).}
\end{figure*}

%% file: figures/bandits_large.pdf_tex
\begingroup%
  \makeatletter%
  \providecommand\color[2][]{%
    \errmessage{(Inkscape) Color is used for the text in Inkscape, but the package 'color.sty' is not loaded}%
    \renewcommand\color[2][]{}%
  }%
  \providecommand\transparent[1]{%
    \errmessage{(Inkscape) Transparency is used (non-zero) for the text in Inkscape, but the package 'transparent.sty' is not loaded}%
    \renewcommand\transparent[1]{}%
  }%
  \providecommand\rotatebox[2]{#2}%
  \newcommand*\fsize{\dimexpr\f@size pt\relax}%
  \newcommand*\lineheight[1]{\fontsize{\fsize}{#1\fsize}\selectfont}%
  \ifx\svgwidth\undefined%
    \setlength{\unitlength}{162.96364371bp}%
    \ifx\svgscale\undefined%
      \relax%
    \else%
      \setlength{\unitlength}{\unitlength * \real{\svgscale}}%
    \fi%
  \else%
    \setlength{\unitlength}{\svgwidth}%
  \fi%
  \global\let\svgwidth\undefined%
  \global\let\svgscale\undefined%
  \makeatother%
  \begin{picture}(1,0.99789783)%
    \lineheight{1}%
    \setlength\tabcolsep{0pt}%
    \put(0,0){\includegraphics[width=\unitlength,page=1]{bandits_large.pdf}}%
    \put(0.92316678,0.78751981){\color[rgb]{0,0,0}\makebox(0,0)[lt]{\lineheight{1.25}\smash{\begin{tabular}[t]{l}1\end{tabular}}}}%
    \put(0.34705505,0.89135285){\color[rgb]{0,0,0}\makebox(0,0)[lt]{\lineheight{1.25}\smash{\begin{tabular}[t]{l}$\bigoplus \boldsymbol{s} = 0$\end{tabular}}}}%
    \put(0.03314624,0.77795558){\color[rgb]{0,0,0}\makebox(0,0)[lt]{\lineheight{1.25}\smash{\begin{tabular}[t]{l}0\end{tabular}}}}%
    \put(0,0){\includegraphics[width=\unitlength,page=2]{bandits_large.pdf}}%
    \put(0.92632843,0.26543913){\color[rgb]{0,0,0}\makebox(0,0)[lt]{\lineheight{1.25}\smash{\begin{tabular}[t]{l}1\end{tabular}}}}%
    \put(0.35021656,0.36927211){\color[rgb]{0,0,0}\makebox(0,0)[lt]{\lineheight{1.25}\smash{\begin{tabular}[t]{l}$\bigoplus \boldsymbol{s} = 1$\end{tabular}}}}%
    \put(0.03630788,0.25587484){\color[rgb]{0,0,0}\makebox(0,0)[lt]{\lineheight{1.25}\smash{\begin{tabular}[t]{l}0\end{tabular}}}}%
  \end{picture}%
\endgroup%

%% file: paper.bbl
\begin{thebibliography}{10}
\providecommand{\url}[1]{#1}
\csname url@samestyle\endcsname
\providecommand{\newblock}{\relax}
\providecommand{\bibinfo}[2]{#2}
\providecommand{\BIBentrySTDinterwordspacing}{\spaceskip=0pt\relax}
\providecommand{\BIBentryALTinterwordstretchfactor}{4}
\providecommand{\BIBentryALTinterwordspacing}{\spaceskip=\fontdimen2\font plus
\BIBentryALTinterwordstretchfactor\fontdimen3\font minus
  \fontdimen4\font\relax}
\providecommand{\BIBforeignlanguage}[2]{{%
\expandafter\ifx\csname l@#1\endcsname\relax
\typeout{** WARNING: IEEEtran.bst: No hyphenation pattern has been}%
\typeout{** loaded for the language `#1'. Using the pattern for}%
\typeout{** the default language instead.}%
\else
\language=\csname l@#1\endcsname
\fi
#2}}
\providecommand{\BIBdecl}{\relax}
\BIBdecl

\bibitem{Benedetti_2019}
M.~Benedetti, E.~Lloyd, S.~Sack, and M.~Fiorentini, ``{P}arameterized quantum
  circuits as machine learning models,'' \emph{Quantum Sci. Technol.}, vol.~4,
  no.~4, p. 043001, 2019.

\bibitem{Liu_2021}
Y.~Liu, S.~Arunachalam, and K.~Temme, ``{A} rigorous and robust quantum
  speed-up in supervised machine learning,'' \emph{Nat. Phys.}, vol.~17, no.~9,
  pp. 1013--1017, 2021.

\bibitem{Sweke_2021}
R.~Sweke, J.-P. Seifert, D.~Hangleiter, and J.~Eisert, ``{O}n the {Q}uantum
  versus {C}lassical {L}earnability of {D}iscrete {D}istributions,''
  \emph{Quantum}, vol.~5, p. 417, 2021.

\bibitem{Sharma_2020}
K.~Sharma, S.~Khatri, M.~Cerezo, and P.~J. Coles, ``{N}oise resilience of
  variational quantum compiling,'' \emph{New J. Phys.}, vol.~22, no.~4, p.
  043006, 2020.

\bibitem{Fontana_2021}
E.~Fontana, N.~Fitzpatrick, D.~M. Ramo, R.~Duncan, and I.~Rungger,
  ``{E}valuating the noise resilience of variational quantum algorithms,''
  \emph{Phys. Rev. A}, vol. 104, no.~2, p. 022403, 2021.

\bibitem{Mitarai_2018}
K.~Mitarai, M.~Negoro, M.~Kitagawa, and K.~Fujii, ``{Q}uantum circuit
  learning,'' \emph{Phys. Rev. A}, vol.~98, no.~3, p. 032309, 2018.

\bibitem{Meyer_2022a}
N.~Meyer, C.~Ufrecht, M.~Periyasamy, D.~D. Scherer, A.~Plinge, and
  C.~Mutschler, ``{A} {S}urvey on {Q}uantum {R}einforcement {L}earning,''
  \emph{arXiv:2211.03464}, 2022.

\bibitem{Sutton_2018}
R.~S. Sutton and A.~G. Barto, \emph{{R}einforcement {L}earning: {A}n
  {I}ntroduction}.\hskip 1em plus 0.5em minus 0.4em\relax MIT press, 2018.

\bibitem{Jerbi_2021}
S.~Jerbi, C.~Gyurik, S.~Marshall, H.~Briegel, and V.~Dunjko, ``{P}arametrized
  {Q}uantum {P}olicies for {R}einforcement {L}earning,'' \emph{Adv. Neural Inf.
  Process. Syst.}, vol.~34, pp. 28\,362--28\,375, 2021.

\bibitem{Lattimore_2013}
T.~Lattimore, M.~Hutter, and P.~Sunehag, ``The sample-complexity of general
  reinforcement learning,'' in \emph{International Conference on Machine
  Learning}, 2013, pp. 28--36.

\bibitem{Amari_1998}
S.-I. Amari, ``Natural {G}radient {W}orks {E}fficiently in {L}earning,''
  \emph{Neural Comput.}, vol.~10, no.~2, pp. 251--276, 1998.

\bibitem{Martens_2020}
J.~Martens, ``{N}ew {I}nsights and {P}erspectives on the {N}atural {G}radient
  {M}ethod,'' \emph{J. Mach. Learn. Res.}, vol.~21, no.~1, pp. 5776--5851,
  2020.

\bibitem{Stokes_2020}
J.~Stokes, J.~Izaac, N.~Killoran, and G.~Carleo, ``{Q}uantum {N}atural
  {G}radient,'' \emph{Quantum}, vol.~4, p. 269, 2020.

\bibitem{Meyer_2023}
N.~Meyer, D.~Scherer, A.~Plinge, C.~Mutschler, and M.~Hartmann, ``Quantum
  policy gradient algorithm with optimized action decoding,'' in
  \emph{International Conference on Machine Learning}, vol. 202.\hskip 1em plus
  0.5em minus 0.4em\relax PMLR, 2023, pp. 24\,592--24\,613.

\bibitem{Jerbi_2022}
S.~Jerbi, A.~Cornelissen, M.~Ozols, and V.~Dunjko, ``Quantum policy gradient
  algorithms,'' \emph{arXiv:2212.09328}, 2022.

\bibitem{Skolik_2023}
A.~Skolik, S.~Mangini, T.~B{\"a}ck, C.~Macchiavello, and V.~Dunjko,
  ``Robustness of quantum reinforcement learning under hardware errors,''
  \emph{EPJ Quantum Technol.}, vol.~10, no.~1, pp. 1--43, 2023.

\bibitem{Kakade_2001}
S.~M. Kakade, ``{A} {N}atural {P}olicy {G}radient,'' \emph{Adv. Neural Inf.
  Process. Syst.}, vol.~14, 2001.

\bibitem{Haug_2021}
T.~Haug and M.~Kim, ``Optimal training of variational quantum algorithms
  without barren plateaus,'' \emph{arXiv:2104.14543}, 2021.

\bibitem{Thanasilp_2021}
S.~Thanasilp, S.~Wang, N.~A. Nghiem, P.~J. Coles, and M.~Cerezo, ``Subtleties
  in the trainability of quantum machine learning models,''
  \emph{arXiv:2110.14753}, 2021.

\bibitem{Yamamoto_2019}
N.~Yamamoto, ``On the natural gradient for variational quantum eigensolver,''
  \emph{arXiv:1909.05074}, 2019.

\bibitem{Sutton_1999}
R.~S. Sutton, D.~McAllester, S.~Singh, and Y.~Mansour, ``{P}olicy {G}radient
  {M}ethods for {R}einforcement {L}earning with {F}unction {A}pproximation,''
  in \emph{Adv. Neural Inf. Process. Syst.}, vol.~12, 1999.

\bibitem{Kullback_1951}
S.~Kullback and R.~A. Leibler, ``On {I}nformation and {S}ufficiency,''
  \emph{Ann. Math. Stat.}, vol.~22, no.~1, pp. 79--86, 1951.

\bibitem{Cheng_2010}
R.~Cheng, ``{Q}uantum {G}eometric {T}ensor ({F}ubini-{S}tudy {M}etric) in
  {S}imple {Q}uantum {S}ystem: {A} pedagogical {I}ntroduction,''
  \emph{arXiv:1012.1337}, 2010.

\bibitem{Perez_2020}
A.~P{\'e}rez-Salinas, A.~Cervera-Lierta, E.~Gil-Fuster, and J.~I. Latorre,
  ``{D}ata re-uploading for a universal quantum classifier,'' \emph{Quantum},
  vol.~4, p. 226, 2020.

\bibitem{Periyasamy_2022}
M.~Periyasamy, N.~Meyer, C.~Ufrecht, D.~D. Scherer, A.~Plinge, and
  C.~Mutschler, ``{I}ncremental {D}ata-{U}ploading for {F}ull-{Q}uantum
  {C}lassification,'' in \emph{IEEE Int. Conf. Quantum Comp. Eng. (QCE)}, 2022,
  pp. 31--37.

\bibitem{Lopatnikova_2021}
A.~Lopatnikova and M.-N. Tran, ``{Q}uantum {S}peedup of {N}atural {G}radient
  for {V}ariational {B}ayes,'' \emph{arXiv:2106.05807}, 2021.

\bibitem{Malago_2013}
L.~Malag{\`o} and M.~Matteucci, ``{R}obust {E}stimation of {N}atural {G}radient
  in {O}ptimization by {R}egularized {L}inear {R}egression,'' in \emph{Geom.
  Sci. Inf., GSI 2013}, 2013, pp. 861--867.

\bibitem{Qiskit_2023}
{Qiskit contributers}, ``{Q}iskit: {A}n {O}pen-source {F}ramework for {Q}uantum
  {C}omputing,'' \url{https://quantum-computing.ibm.com/}, 2023.

\bibitem{Cultrera_2020}
A.~Cultrera and L.~Callegaro, ``A simple algorithm to find the l-curve corner
  in the regularisation of ill-posed inverse problems,'' \emph{IOP SciNotes},
  vol.~1, no.~2, p. 025004, 2020.

\bibitem{Spall_1998}
J.~C. Spall, ``{A}n {O}verview of the {S}imultaneous {P}erturbation {M}ethod
  for {E}fficient {O}ptimization,'' \emph{Johns Hopkins APL Tech. Dig.},
  vol.~19, no.~4, pp. 482--492, 1998.

\bibitem{Gacon_2021}
J.~Gacon, C.~Zoufal, G.~Carleo, and S.~Woerner, ``{S}imultaneous {P}erturbation
  {S}tochastic {A}pproximation of the {Q}uantum {F}isher {I}nformation,''
  \emph{Quantum}, vol.~5, p. 567, 2021.

\bibitem{Franz_2022}
M.~Franz, L.~Wolf, M.~Periyasamy, C.~Uftrecht, D.~D. Scherer, A.~Plinge,
  C.~Mutschler, and W.~Mauerer, ``{U}ncovering instabilities in
  variational-quantum deep {Q}-networks,'' \emph{J. Franklin Inst.}, 2022.

\bibitem{IBMquantum_2021}
{IBM Quantum}, ``{Q}iskit {R}untime {S}ervice, {S}ampler primitive (version
  0.9.1),'' \url{https://quantum-computing.ibm.com/}, 2023.

\bibitem{Wiedmann_2023}
M.~Wiedmann, M.~H{\"o}lle, M.~Periyasamy, N.~Meyer, C.~Ufrecht, D.~D. Scherer,
  A.~Plinge, and C.~Mutschler, ``An empirical comparison of optimizers for
  quantum machine learning with spsa-based gradients,''
  \emph{arXiv:2305.00224}, 2023.

\bibitem{Nation_2021}
P.~D. Nation, H.~Kang, N.~Sundaresan, and J.~M. Gambetta, ``Scalable mitigation
  of measurement errors on quantum computers,'' \emph{PRX Quantum}, vol.~2,
  no.~4, p. 040326, 2021.

\bibitem{Wang_2021a}
S.~Wang, E.~Fontana, M.~Cerezo, K.~Sharma, A.~Sone, L.~Cincio, and P.~J. Coles,
  ``Noise-induced barren plateaus in variational quantum algorithms,''
  \emph{Nature communications}, vol.~12, no.~1, p. 6961, 2021.

\bibitem{Wang_2021b}
S.~Wang, P.~Czarnik, A.~Arrasmith, M.~Cerezo, L.~Cincio, and P.~J. Coles, ``Can
  error mitigation improve trainability of noisy variational quantum
  algorithms?'' \emph{arXiv:2109.01051}, 2021.

\bibitem{Moll_2018}
N.~Moll, P.~Barkoutsos, L.~S. Bishop, J.~M. Chow, A.~Cross, D.~J. Egger,
  S.~Filipp, A.~Fuhrer, J.~M. Gambetta, M.~Ganzhorn \emph{et~al.}, ``{Q}uantum
  optimization using variational algorithms on near-term quantum devices,''
  \emph{Quantum Sci. Technol.}, vol.~3, no.~3, p. 030503, 2018.

\end{thebibliography}
